\documentclass[thmsa,letterpaper,12pt]{article}
\usepackage{sw20lart}

\input tcilatex
\QQQ{Language}{
American English
}

\begin{document}

\title{\textbf{Nonlinear Fractional Dynamics on a Lattice with Long Range
Interactions}}
\author{N. Laskin and G. Zaslavsky \\
Courant Institute of Mathematical Sciences\\
New York University\\
251 Mercer Street, NY 10012}
\date{}
\maketitle

\begin{abstract}
A unified approach has been developed to study nonlinear dynamics of a 1D
lattice of particles with long-range power-law interaction. A classical case
is treated in the framework of the generalization of the well-known
Frenkel-Kontorova chain model for the non-nearest interactions. Quantum
dynamics is considered following Davydov's approach for molecular excitons.

In the continuum limit the problem is reduced to dynamical equations with
fractional derivatives resulting from the fractional power of the long-range
interaction. Fractional generalizations of the sine-Gordon, nonlinear
Schr\"odinger, and Hilbert-Schr\"odinger equations have been found.

There exists a critical value of the power $s$ of the long-range potential.
Below the critical value ($s<3$, $s\neq 1$, $2$) we obtain equations with
fractional derivatives while for $s\geq 3$ we have the well-known nonlinear
dynamical equations with space derivatives of integer order.

Long-range interaction impact on the quantum lattice propagator has been
studied. We have shown that the quantum exciton propagator exhibits
transition from the well-known Gaussian-like behavior to a power-law decay
due to the long-range interaction. A link between 1D quantum lattice
dynamics in the imaginary time domain and a random walk model has been
discussed.

\textit{PACS }number(s): 03.65.-w, 03.65. Db, 05.30.-d, 05.40. Fb

\end{abstract}

\section{Introduction}

Discrete nonlinear lattice models have become one of the most popular models
of nonlinear physics, often used to investigate a rather broad set of
physical phenomena and systems. The first simplest 1D lattice model was
introduced a long time ago by Frenkel and Kontorova \cite{Frenkel} to study
the structure and dynamics of dislocations in metals (for more details see
review \cite{Kivshar} and references therein). The Frenkel and Kontorova
discrete classical dynamics is described as a 1D chain of atoms with the
nearest-neighbor intersite/interatomic interaction and a periodic on-site
potential. Despite its simplicity the model has many applications in
solid-state and nonlinear physics, including propagation of charge-density
waves, the dynamics of absorbed layers of atoms on crystal surfaces,
commensurable-incommensurable phase transitions, and domain walls in
magnetically ordered structures, see for details \cite{Kivshar}. Moreover
the model is surprisingly attractive because it provides exactly integrable
cases in discrete and continuous approximations including the well-known
sine-Gordon nonlinear equation (see, for example \cite{Zakharov}).

A specific, but not less important application, is the case of long-range
interatomic interaction, which is the main subject of this paper and which
has been studied for different physical systems. Examples include
ferromagnetic chains \cite{Nakano}, exciton transfer in molecular crystals 
\cite{Davydov}, \cite{Scott}, commensurable-incommensurable phase
transitions \cite{Pokrovsky}, theory of Josephson junctions \cite{Jos1}, 
\cite{Alfimov}. To model the long-range interaction potential in 1D lattice,
few potentials have been applied: power-law interaction \cite{Pokrovsky}, 
\cite{Kosevich}, Lennard-Jones long-range coupling \cite{Ishimori} and
exponential interaction of Kac-Beker form \cite{Kac}.

Since the Fermi, Pasta and Ulam work \cite{Fermi} analytical developments
and numerical simulation on discrete nonlinear dynamical equations have
provided numerous types of nonlinear excitations with various properties:
solitons, continuous and discrete breathers \cite{Kivshar}, \cite{Flach1}, 
\cite{Flach}, self-trapping states \cite{Tuszyn} and others.

Another broad area of applications is related to the coupled system of
oscillators with long-range interactions and the phenomenon of
synchronization in the system \cite{Z1}-\cite{Z3}. In many situations,
impact of long-range coupling can be compared to a phase transition \cite{Z4}%
-\cite{Z6}. Long-range coupling can be presented as nonlocality with finite
space scale $a$, when the coupling energy is proportional to $\exp
(-|x_n-x_m|/a)$ for atoms located at the points $x_n$ and $x_m$, or with
scale free interaction when the coupling energy is proportional to $%
|x_n-x_m|^{-s}$ ($s>0$). In addition to the well known interaction with
integer values of $s$ , some complex media can be described by fractional
values of $s$ (see, for example, references in \cite{Z7}).

The goal of this paper is to focus on analytical developments on classical
and quantum 1D lattice dynamics with fractional power-law long-range
interaction. The long-range interaction leads in general to a nonlocal
integral term in the equation of motion if we go from discrete to continues
space. We show that in some particular cases for power-law interaction with
non-integer power $s$ the integral term can be expressed through a
fractional order derivative. In other words, nonlocality originating from
the long-range interaction is revealed in the dynamics in the form of space
derivatives of fractional order. The appearance of fractional differential
equations in the continuum space limit of lattice dynamics leads to a
possibility to apply powerful tools of random processes theory \cite{Z9}, 
\cite{Z13} and fractional calculus \cite{Z10}-\cite{Z14} to the lattice
dynamics and kinetics. Particularly, it helps to obtain some qualitative
results based on the known features of fractional derivatives. As an
example, let us mention that the fractional oscillator appears to have power
law dissipation at $t\rightarrow \infty $ and, being perturbed, gives rise
to a new type of stochastic attractor \cite{Z7}. In the case of a fractional
space derivative, one can expect new types of synchronizations and coherent
structures \cite{Z8}, where our approach to obtain equations with fractional
derivative is applied to discrete model of oscillating media with long-range
power-type coupling between oscillators. It also worthwhile to mention that
fractional dynamical equations prove to be an adequate approach to study
chaotic dynamical systems and their anomalous transport properties \cite
{Zaslavsky1}.

We present new developments on a 1D lattice model with fractional power-law
interatomic interaction defined by fractional values of the parameter $s$
and non-linear self-interaction potential for classical and quantum
mechanical consideration. We show that the dynamics of the 1D lattice can be
equivalently presented by the corresponding fractional nonlinear equation in
the long-wave limit. We concentrate on the conditions of such equivalence,
type of the equations with fractional derivatives and some related
properties. As example of our developments fractional sine-Gordon and
wave-Hilbert nonlinear equations have been found for classical lattice
dynamics, and fractional nonlinear Schr\"odinger and nonlinear
Hilbert-Schrodinger equations have been obtained for quantum lattice
dynamics in the long wave limit.

The paper is organized as follows. In Sec.2 we study classical 1D lattice
non-linear dynamics. It has been found that depending on the long-range
interaction parameter $s$, the interaction term can be presented in
long-wave limit in differential or fractional differential form which
results either as a nonlinear wave equation or as a nonlinear fractional
wave equation. Examples are fractional sine-Gordon and wave-Hilbert
nonlinear equations.

Sec.3 focuses on quantum 1D lattice dynamics. Depending on the long-range
interaction parameter $s$, the interaction term can be presented in the
long-wave limit in differential or fractional differential form which
results in the nonlinear Schr\"odinger equation or in a nonlinear
Hilbert-Schr\"odinger equation.

Quantum lattice propagator is introduced in Sec.4. It has been observed that
depending on the parameter $s$ the quantum lattice propagator exhibits a
transition from the well-known Gaussian-like behavior to power-law decay due
to long-range interaction. We show a link between the quantum lattice
dynamics and random walk in the imaginary time domain.

In conclusion we outline the new developments and discuss their
applicability to other fields.

In Appendices we briefly review the definition of the polylogarithm and its
integral and power series representations.

\section{Classical nonlinear lattice dynamics}

Nonlinear classical discrete dynamics of a 1D infinite lattice can be
described by Hamiltonian function $H(u)$, which depends on particle
(''atom'') displacement $u_n(t)$ at the site $n$

\begin{equation}
H(u)=\frac 12M\sum\limits_{n=-\infty }^\infty \dot u_n^2+\frac
12\sum\limits_{n=-\infty }^\infty \sum\limits\Sb m=-\infty  \\ n\neq m 
\endSb ^\infty V_{n-m}(u_n-u_m)^2+\sum\limits_{n=-\infty }^\infty U(u_n),
\label{eq1C}
\end{equation}
where the first term is the kinetic energy of the chain, $M$ is particle
mass, and $V_{n-m}$ is interaction matrix which describes long-range
interaction between sites $n$ and $m$

\begin{equation}
V_{n-m}=\frac{V_0}{|n-m|^s}.  \label{eq2C}
\end{equation}
Here $V_0$ is interaction constant, parameter $s$ covers different physical
models, and $U(u)$ characterizes nonlinear interaction of the lattice atoms
with external on-site potential. Integer values of $s$ can be used to
describe the well-known physical models: the nearest-neighbor approximation (%
$s=\infty $), the dipole-dipole interaction ($s=3$), the Coulomb potential ($%
s=1$).

Our main interest will be in fractional values of $s$ that can appear for
more sophisticated interaction potentials attributed to complex media. The
equation of motion follows from Eq.(\ref{eq1C})

\begin{equation}
M\ddot u_n+\sum\limits\Sb m=-\infty  \\ n\neq m  \endSb ^\infty
V_{n-m}(u_n-u_m)+\frac{\partial U(u)}{\partial u_n}=0,  \label{eq3C}
\end{equation}
To go from the discrete version Eq.(\ref{eq3C}) to the continuum one, let us
define 
\begin{equation}
v(k,t)=\sum\limits_{n=-\infty }^\infty e^{-ikn}u_n(t),  \label{eq4C}
\end{equation}
where $u_n(t)$ is related to $v(k,t)$ as follows

\begin{equation}
u_n(t)=\frac 1{2\pi }\int\limits_{-\pi }^\pi dke^{ikn}v(k,t),  \label{eq5C}
\end{equation}
and $k$ can be considered as a wave number.

Then the second term in Eq.(\ref{eq1C}) reads

\begin{equation}
\frac 12\sum\limits\Sb n,m  \\ n\neq m  \endSb V_{n-m}(u_n-u_m)^2=\frac
1{2\pi }\int\limits_{-\pi }^\pi dk\mathcal{V}(k)|v(k,t)|^2,  \label{eq6C}
\end{equation}
where function $\mathcal{V}(k)$ has been defined by

\begin{equation}
\mathcal{V}(k)=V(0)-V(k),  \label{eq7C}
\end{equation}
and $V(k)$ is

\begin{equation}
V(k)=\sum\limits\Sb n=-\infty  \\ n\neq 0  \endSb ^\infty e^{-ikn}V_n.
\label{eq8C}
\end{equation}

In the long wave limit when the wavelength exceeds the intersite scale we
may consider $v(k,t)$ as a $k$th Fourier component of function $u(x,t)$, $%
u_n(t)\stackunder{k\longrightarrow 0}{\longrightarrow }$ $u(x,t)$. That is
the functions $v(k,t)$ and $u(x,t)$ are related each other by the Fourier
transform

\begin{equation}
u(x,t)=\frac 1{2\pi }\int\limits_{-\infty }^\infty dke^{ikx}v(k,t),\qquad
\qquad \qquad v(k,t)=\int\limits_{-\infty }^\infty dxe^{-ikx}u(x,t).
\label{eq10C}
\end{equation}
Therefore, from Eq.(\ref{eq6C}) we have

\begin{equation}
\frac 12\sum\limits_{n=-\infty }^\infty \sum\limits\Sb m=-\infty  \\ n\neq m 
\endSb ^\infty V_{n-m}(u_n-u_m)^2=\frac 12\int\limits_{-\infty }^\infty
dx\int\limits_{-\infty }^\infty dyu(x,t)W(x-y)u(y,t),  \label{eq11C}
\end{equation}
with the kernel

\begin{equation}
W(x)=\frac 1\pi \int\limits_{-\infty }^\infty dke^{ikx}\mathcal{V}(k).
\label{eq111C}
\end{equation}
that comes from the long-range interaction matrix $V_{n-m}$ and it possesses
the properties:

(i) $W(x)$ is an even function;

(ii) $\int\limits_{-\infty }^\infty dxW(x)=0.$

One can express Eq.(\ref{eq11C}) in a non-local kinematic form

\begin{equation}
\frac 12\sum\limits\Sb n,m  \\ n\neq m  \endSb J_{n-m}(u_n-u_m)^2=\frac
12\int\limits_{-\infty }^\infty dx\int\limits_{-\infty }^\infty dy(\partial
_xu(x,t))K(x-y)\partial _yu(y,t),  \label{eq112C}
\end{equation}
where notation $\partial _x=\partial /\partial x$ has been introduced and
the relationship between kernel $K(x)$ and function $\mathcal{V}(k)$ (see
Eq.(\ref{eq7C})) is given by

\begin{equation}
K(x)=\frac 1\pi \int\limits_{-\infty }^\infty dke^{ikx}\frac{\mathcal{V}(k)}{%
k^2}.  \label{eq12C}
\end{equation}
In other words, the kernels $W(x)$ and $K(x)$ are related to each other as

\begin{equation}
W(x)=-\partial _x^2K(x).  \label{eq122C}
\end{equation}
Thus, in continuum limit the equation of motion becomes 
\begin{equation}
M\ddot u(x,t)-\int\limits_{-\infty }^\infty dy\partial _xK(x-y)\partial
_yu(y,t)+\frac{\partial U(u)}{\partial u(x,t)}=0,  \label{eq13C}
\end{equation}
or in a symbolic form

\begin{equation}
M\ddot u(x,t)-\partial _x(\widehat{K}\partial _xu)+\frac{\partial U(u)}{%
\partial u(x,t)}=0,  \label{eq14C}
\end{equation}
and the operator $\widehat{K}$, as it follows from Eq.(\ref{eq12C}), can be
considered as operator of multiplication in wave number (momentum) space

\begin{equation}
\widehat{K}u(x,t)(k)=\frac{\mathcal{V}(k)}{k^2}v(k,t).  \label{eq15C}
\end{equation}
Equation (\ref{eq13C}) is integro-differential equation of motion. The
integral part comes from the long-range interaction term Eq.(\ref{eq112C}).
To get the differential equation of motion we use the properties of function 
$\mathcal{V}(k)$ at $k\rightarrow 0$, which can be obtained from the
asymptotics of polylogarithm (we provide the properties of the polylogarithm
function and the expression for $\mathcal{V}(k)$ in the Appendices),

\begin{equation}
\mathcal{V}(k)\sim \frac{\pi V_0}{\Gamma (s)\sin (\frac{\pi (s-1)}2)}%
|k|^{s-1}=D_s|k|^{s-1},\quad \quad 2\leq s<3,  \label{eqA15C}
\end{equation}

\begin{equation}
\mathcal{V}(k)\sim -\frac{V_0k^2}2\ln k^2,\quad s=3,  \label{eqA16C}
\end{equation}

\begin{equation}
\mathcal{V}(k)\sim \frac{V_0\zeta (s-2)}2k^2,\quad s>3,  \label{eqA17C}
\end{equation}
where $\Gamma (s)$ is $\Gamma $-function, $\zeta (s)$ is Riemann zeta
function and coefficient $D_s$ is defined by 
\begin{equation}
D_s=\frac{\pi V_0}{\Gamma (s)\sin (\frac{\pi (s-1)}2)}.  \label{eqA155C}
\end{equation}
It is seen from Eq.(\ref{eqA15C}) that fractional powers of $k$ occurs for
the interactions with $2\leq s<3$ only.

Going back to the coordinate space yields for fractional powers of $|k|$
fractional Riesz derivative of order $\alpha $ \cite{Z13}, \cite{Z10}

\begin{equation}
\partial _x^\alpha u(x,t)=-\frac 1{2\cos (\pi \alpha /2)}\left( \frac{%
d^\alpha }{dx^\alpha }+\frac{d^\alpha }{d(-x)^\alpha }\right) u(x,t),\qquad
1<\alpha \leq 2,  \label{eqB15C}
\end{equation}
where $d^\alpha /d(\pm x)^\alpha $ are the Riemann-Liouville derivatives

\[
\frac{d^\alpha }{dx^\alpha }f(x)=\frac 1{\Gamma (n-\alpha )}\frac{\partial ^n%
}{\partial x^n}\int\limits_{-\infty }^x\frac{dyf(y)}{(x-y)^{\alpha -n+1}}, 
\]

\[
\frac{d^\alpha }{d(-x)^\alpha }f(x)=\frac 1{\Gamma (n-\alpha )}\frac{%
\partial ^n}{\partial x^n}\int\limits_x^\infty \frac{dyf(y)}{(y-x)^{\alpha
-n+1}}, 
\]
with $n-1<\alpha <n$ and integer $n$. Substitution of Eq.(\ref{eqA15C}) into
Eq.(\ref{eq13C}) yields a fractional nonlinear wave equation

\begin{equation}
M\ddot u(x,t)-D_s\partial _x^{s-1}u(x,t)+\frac{\partial U(u)}{\partial u(x,t)%
}=0,  \label{eq28C}
\end{equation}
where the Riesz fractional derivative has been transformed into \cite{Z13}, 
\cite{Laskin1}

\begin{equation}
\partial _x^{s-1}u(x,t)=-\frac 1{2\pi }\int\limits_{-\infty }^\infty
dke^{ikx}|k|^{s-1}v(k,t),  \label{eq29C}
\end{equation}
and coefficient $D_s$ is defined by Eq.(\ref{eqA155C}).

When the nonlinear external potential $U(u)=U_0(1-\cos u),$ we get the
fractional sine-Gordon equation \cite{FracGordon} 
\begin{equation}
M\ddot u(x,t)-D_s\partial _x^{s-1}u(x,t)+U_0\sin u(x,t)=0.  \label{eq31C}
\end{equation}
To get differences for the phonon spectrum of fractional and standard
sine-Gordon equations let's consider a linear equation of the form

\begin{equation}
\ddot u(x,t)-c_0^2\partial _{xx}^2u(x,t)+m^2u(x,t)=0,  \label{eqL1}
\end{equation}
where $c_0$ is velocity of wave and $m$ is the particle mass. Plane wave
solution of this equation has the form

\begin{equation}
u(x,t)=Ae^{i(\omega t+kx)},  \label{eqL2}
\end{equation}
with constant $A$, frequency $\omega $, and wave number $k$. Then the
dispersion law is

\begin{equation}
\omega (k)=\pm \sqrt{m^2+c_0^2k^2}.  \label{eql3}
\end{equation}
For the fractional linear wave equation

\begin{equation}
\ddot u(x,t)-b^{s-3}c_0^2\partial _x^{s-1}u(x,t)+m^2u(x,t)=0,  \label{eqL4}
\end{equation}
where $\partial _x^{s-1}$ is defined by Eq.(\ref{eq29C}), $b$ is the scale
constant attributed to the fractional wave equation, $c_0$ is wave velocity,
the dispersion law is

\begin{equation}
\omega (k)=\pm \sqrt{m^2+b^{s-3}c_0^2|k|^{s-1}},\qquad 2<s<3.  \label{eqL5}
\end{equation}

For fairly small values of $m$ and not too small $k$ we obtain

\begin{equation}
\omega (k)\approx b^{(s-3)/2}c_0|k|^{(s-1)/2},  \label{eqL6}
\end{equation}
which results in phase $\mathrm{v}_{ph}$ and group $\mathrm{v}_g$ velocities
equal

\begin{equation}
\mathrm{v}_{ph}=\frac{\omega (k)}k=b^{(s-3)/2}c_0/|k|^{(3-s)/2},\qquad 2<s<3,
\label{eqL7}
\end{equation}
\begin{equation}
\mathrm{v}_g=\frac{\partial \omega (|k|)}{\partial |k|}=\frac{s-1}%
2b^{(s-3)/2}c_0/|k|^{(3-s)/2}=\frac{s-1}2\mathrm{v}_{ph},\qquad 2<s<3.
\label{eqL77}
\end{equation}
These expressions tend to infinity for $k\rightarrow 0$, and we arrive at
new physical properties of the lattice of particles just because of
fractality of the long-range interaction. For small enough $k$ the mass $m$
can not be neglected if $m\neq 0$. Several other examples of equations can
be easily obtained using the general scheme of Eq.(\ref{eq28C})

It follows from Eq.(\ref{eqA15C}) that in the case $s=2$ the function $%
\mathcal{V}(k)$ takes the form

\begin{equation}
\mathcal{V}(k)=\pi V_0|k|.  \label{eq33C}
\end{equation}
Substitution of this expression into Eqs.(\ref{eq12C}) and (\ref{eq14C})
yields a nonlinear wave-Hilbert equation

\begin{equation}
M\ddot u(x,t)-\pi V_0\mathcal{H}\left\{ \partial _xu(x,t)\right\} +\frac{%
\partial U(u)}{\partial u(x,t)}=0,  \label{eq34C}
\end{equation}
where $\mathcal{H}$ is the Hilbert integral transform defined by

\begin{equation}
\mathcal{H}\{\phi (x)\}=\mathrm{P}\int\limits_{-\infty }^\infty dy\frac{\phi
(y)}{y-x},  \label{eq35C}
\end{equation}
and $\mathrm{P}$ stands for the Cauchy principal value of the integral.

When $U(u)=U_0(1-\cos u)$ we find the sine-Hilbert equation \cite{Gurevich}, 
\cite{Silin} 
\begin{equation}
M\ddot u(x,t)-\pi V_0\mathcal{H}\left\{ \partial _xu(x,t)\right\} +U_0\sin
u(x,t)=0.  \label{eq36C}
\end{equation}

Examples of physical problems with nonlocal interaction like $\mathcal{V}(k)$
given by Eq.(\ref{eq33C}) can be found in nonlocal Josephson electrodynamics 
\cite{Joseph1} where Eq.(\ref{eq36C}) is one of the basic model equations.
Let's note that some exact solutions of this equation have been found and
their stability under weak perturbations has been analyzed \cite{Silin}-\cite
{Joseph2}.

Finally, when $s>3$ and $U(u)=U_0(1-\cos u)$ we get the sine-Gordon equation 
\cite{Zakharov},

\begin{equation}
M\ddot u(x,t)-\frac{V_0\zeta (s-2)}2\partial _{xx}^2u(x,t)+U_0\sin u(x,t)=0,
\label{eq37C}
\end{equation}
where $\sqrt{\frac{V_0\zeta (s-2)}{2M}}$ can be interpreted as the wave
velocity.

\section{Quantum lattice dynamics}

\subsection{Lattice Hamiltonian with long-range interaction}

To model a 1D quantum lattice dynamics we follow \cite{Davydov}, \cite{Scott}
and consider a linear, rigid arrangement of sites with one molecule at each
lattice site. To describe creation (annihilation) of molecular excitation
or, for simplicity exciton, at the site $n$ we introduce exciton creation $%
b_n^{+}$ and annihilation $b_n$ operators. Operators $b_n^{+}$ and $b_n$
satisfy the commutation relations $[b_n^{+},b_m]=\delta _{n,m}$, $%
[b_n,b_m]=0 $, $[b_n^{+},b_m^{+}]=0$ that is $b_n^{+}$ and $b_n$ are the
Bose operators.

With the help of the operators $b_n^{+}$ and $b_n$ the well-known
Hamiltonian operator of excitons is expressed as \cite{Davydov}, \cite{Scott}%
, \cite{Agranovich},

\begin{equation}
\widehat{H}_D=\varepsilon \sum\limits_{n=-\infty }^\infty
b_n^{+}b_n-J\sum\limits_{n=-\infty }^\infty
(b_n^{+}b_{n+1}+b_n^{+}b_{n-1})=\varepsilon \sum\limits_{n=-\infty }^\infty
b_n^{+}b_n-\sum\limits_{n,m=-\infty }^\infty J_{n,m}b_n^{+}b_m,  \label{eq0q}
\end{equation}
where $\varepsilon $ is constant exciton energy, $J$ is interaction constant
and the excitation transfer matrix element $J_{n,m}$ between sites $n$ and $%
m $ can be written as 
\begin{equation}
J_{n,m}=J(\delta _{\,(n+1),\,m}+\delta _{\,(n-1),\,m}),  \label{eq1q}
\end{equation}
the Kroneker symbols $\delta _{m,\,n\pm 1}$ mean that only the
nearest-neighbor interaction had been considered. The interaction term $%
J_{n,m}b_n^{+}b_m$ in Eq.(\ref{eq0q}) is responsible for exciton transfer
from site $n$ to the nearest neighbor sites $n\pm 1$.

To go beyond the nearest-neighbor interaction we introduce the long-range
excitation transfer matrix $J_{n,m}^{LR}$ which describes power-law
interaction between sites $n$ and $m$ (see Eq.(\ref{eq2C}))

\begin{equation}
J_{n,m}^{LR}\equiv V_{n-m}=\frac{V_0}{|n-m|^s},\qquad n\neq m.  \label{eq2q}
\end{equation}
Then we can rewrite the Hamiltonian operator $\widehat{H}_{LR}$ with
long-range excitation transfer,

\begin{equation}
\widehat{H}_{LR}=\varepsilon \sum\limits_nb_n^{+}b_n-\sum\limits\Sb n,m  \\ %
n\neq m  \endSb J_{n,m}^{LR}b_n^{+}b_m.  \label{eq3q}
\end{equation}

Follow Davydov's ansatz \cite{Davydov}, \cite{Scott} we define the
eigenstate $|\phi (t)>$ of a quantum exciton as a superposition:

\begin{equation}
|\phi (t)>=\sum\limits_n\psi _n(t)b_n^{+}|0>,  \label{eq4q}
\end{equation}
where$|0>$ is the vacuum state of exciton system and $\psi _n(t)$ is the
exciton wave function. In the $|\phi (t)>$ basis the Hamiltonian $\widehat{H}%
_{LR}$ becomes the Hamiltonian function $H_{LR}(\psi ,\psi ^{*})$ 
\begin{equation}
H_{LR}(\psi ,\psi ^{*})=<\phi (t)|\widehat{H}_{LR}|\phi (t)>=\varepsilon
\sum\limits_n\psi _n^{*}(t)\psi _n(t)-\sum\limits\Sb n,m  \\ n\neq m  \endSb %
V_{n-m}\psi _n^{*}(t)\psi _m(t),  \label{eq5q}
\end{equation}
where $V_{n-m}$ is given by Eq.(\ref{eq2q}). Considering nonlinearity in
addition to the long-range interaction, we add a nonlinear term $U(|\psi |)$
to the right hand of Eq.(\ref{eq5q})$,$ and, thus, we obtain for the
Hamiltonian $H(\psi ,\psi ^{*})$ of 1D lattice 
\begin{equation}
H(\psi ,\psi ^{*})=\varepsilon \sum\limits_n\psi _n^{*}(t)\psi
_n(t)-\sum\limits\Sb n,m  \\ n\neq m  \endSb V_{n-m}\psi _n^{*}(t)\psi
_m(t)+\sum\limits_nU(|\psi _n|).  \label{eq6q}
\end{equation}

The nonlinear Schr\"odinger equation in discrete space $i\hbar \partial \psi
_n(t)/\partial t=\delta H(\psi )/\delta \psi _n^{*}$ generated by the above
Hamiltonian has the form

\begin{equation}
i\hbar \frac{\partial \psi _n(t)}{\partial t}=\varepsilon \psi
_n(t)-\sum\limits\Sb m  \\ (n\neq m)  \endSb V_{n-m}\psi _m(t)+\frac{\delta
U(|\psi |)}{\delta \psi _n^{*}},  \label{eq5}
\end{equation}
where $\hbar $ is the Planck constant.

To get continuum space equation let us follow the same approach as in Sec.2
and introduce wave function $\varphi (k,t)$,

\begin{equation}
\varphi (k,t)=\sum\limits_{n=-\infty }^\infty e^{-ikn}\psi _n(t),
\label{eq6}
\end{equation}
where $\psi _n(t)$ is related to $\varphi (k,t)$ as

\begin{equation}
\psi _n(t)=\frac 1{2\pi }\int\limits_{-\pi }^\pi dke^{ikn}\varphi (k,t),
\label{eq7}
\end{equation}
and $k$ can be considered as a wave number. Substitution of Eq.(\ref{eq7})
into Eq.(\ref{eq5q}) yields

\begin{equation}
H_{LR}=\frac 1{2\pi }\int\limits_{-\pi }^\pi dk(\omega +\mathcal{V}%
(k))|\varphi (k,t)|^2,  \label{eq7A}
\end{equation}
with the function $\mathcal{V}(k)$ given by Eq.(\ref{eq7C}) and the energy
parameter $\omega $ which is 
\begin{equation}
\omega =\varepsilon -V(0),  \label{eq8q}
\end{equation}
here $V(0)=\sum\limits_{n=-\infty }^\infty V_n$ (see the definition given by
Eq.(\ref{eq8C})).

At this point we may consider $\varphi (k,t)$ as the wave number
representation of continuous space wave function $\psi (x,t)$. That is the
functions $\varphi (k,t)$ and $\psi (x,t)$ are related to each other by the
Fourier transform

\begin{equation}
\psi (x,t)=\frac 1{2\pi }\int\limits_{-\infty }^\infty dke^{ikx}\varphi
(k,t),\qquad \qquad \varphi (k,t)=\int\limits_{-\infty }^\infty
dxe^{-ikx}\psi (x,t).  \label{eq10}
\end{equation}

Therefore, we have for $H_{LR}$ in coordinate space

\begin{equation}
H_{LR}(\psi )=\omega \int\limits_{-\infty }^\infty dx|\psi (x,t)|^2+\frac
12\int\limits_{-\infty }^\infty dx\int\limits_{-\infty }^\infty dy\psi
^{*}(x,t)W(x-y)\psi (y,t),  \label{eq11}
\end{equation}
where the kernel $W(x)$ is defined by Eq.(\ref{eq111C}). To get kinematic
form of the second term in the right side of Eq.(\ref{eq11}) we follow Eqs.(%
\ref{eq12C}) and (\ref{eq122C}) and find,

\begin{equation}
H_{LR}(\psi )=\omega \int\limits_{-\infty }^\infty dx|\psi (x,t)|^2+\frac
12\int\limits_{-\infty }^\infty dx\int\limits_{-\infty }^\infty dy(\partial
_x\psi ^{*}(x,t))K(x-y)\partial _y\psi (y,t),  \label{eq1111}
\end{equation}
where the kernel $K(x)$ is defined by Eq.(\ref{eq12C}).

Further, by adding a nonlinear term to right hand of (\ref{eq1111}) we
obtain the integro-differential (non local) nonlinear Schr\"odinger type
equation which can written in the following compact way

\begin{equation}
i\hbar \frac{\partial \psi (x,t)}{\partial t}=\omega \psi (x,t)+\partial _x(%
\widehat{K}\partial _x\psi )+\frac{\delta U(|\psi |)}{\delta \psi ^{*}(x,t)},
\label{eq13}
\end{equation}
where the operator $\widehat{K}$ acts on wave function $\psi (x,t)$ as

\begin{equation}
\widehat{K}\psi (x,t)=\frac 1{2\pi }\int\limits_{-\infty }^\infty dke^{ikx}%
\frac{\mathcal{V}(k)}{k^2}\varphi (k,t),  \label{eq14}
\end{equation}
or in wave number (momentum) representation

\begin{equation}
\widehat{K}\psi (x,t)(k)=\frac{\mathcal{V}(k)}{k^2}\varphi (k,t).
\label{eq15}
\end{equation}
Using the properties of function $\mathcal{V}(k)$ in the limit $k\rightarrow
0$ (see Eqs.(\ref{eqA15C})-(\ref{eqA17C})) we can get different special
forms of general Eq.(\ref{eq13}).

\subsection{Fractional nonlinear Schr\"odinger equation}

Substitution Eq.(\ref{eqA15C}) into Eqs.(\ref{eq13}) and (\ref{eq14}) yields
a new nonlinear fractional differential equation

\begin{equation}
i\hbar \frac{\partial \psi (x,t)}{\partial t}=\omega \psi (x,t)-D_s\partial
_x^{s-1}\psi (x,t)+\frac{\delta U(|\psi |)}{\delta \psi ^{*}(x,t)},\qquad
2\leq s<3,  \label{eq25}
\end{equation}
where we use the definition of the Riesz fractional derivative \cite{Z13}, 
\cite{Z10}

\begin{equation}
\partial _x^{s-1}\psi (x,t)=-\frac 1{2\pi }\int\limits_{-\infty }^\infty
dke^{ikx}|k|^{s-1}\varphi (k,t).  \label{eq26}
\end{equation}
In Eq.(\ref{eq25}) coefficient $D_s$ is given by Eq.(\ref{eqA155C}) and $%
\omega $ is the constant defined by Eq.(\ref{eq8q}). For the case when

\begin{equation}
U(|\psi |)=U(x)|\psi |^2,  \label{eq28L}
\end{equation}
Eq.(\ref{eq25}) is reduced to the linear fractional Schr\"odinger equation 
\cite{Laskin1}, \cite{Laskin2}, \cite{Laskin3} 
\begin{equation}
i\hbar \frac{\partial \phi (x,t)}{\partial t}=-D_s\partial _x^{s-1}\phi
(x,t)+U(x)\phi (x,t),  \label{eq29}
\end{equation}

where the wave function $\phi (x,t)$ is related to the wave function $\psi
(x,t)$ by

\begin{equation}
\phi (x,t)=\exp \{i\frac{\omega t}\hbar \}\psi (x,t).  \label{eq28QL}
\end{equation}

Generalization of Eq.(\ref{eq29}) for 3D space and the applications of the
1D and 3D linear fractional Schr\"odinger equation to quantum mechanical
problems have been developed in \cite{Laskin3}. In these papers, a particle
in infinite potential well, fractional oscillator, and fractional Bohr atom
have been studied and the energy spectra for these three quantum mechanical
problems have been obtained using Eq.(\ref{eq29}) and its 3D generalization.
New physical issues following from quantum mechanical applications of Eq.(%
\ref{eq29}) have been discussed in \cite{Laskin3}.

For nonlinearity of the form

\begin{equation}
U(|\psi |)=\frac{U_0}2|\psi |^4,  \label{eq29L}
\end{equation}
with a constant $U_0$, Eq.(\ref{eq25}) is reduced to the fractional
nonlinear Schr\"odinger equation \cite{Z15} - \cite{Z17} 
\begin{equation}
i\hbar \frac{\partial \phi (x,t)}{\partial t}=-D_s\partial _x^{s-1}\phi
(x,t)+U_0|\phi (x,t)|^2\phi (x,t),  \label{eq30}
\end{equation}
where $\phi (x,t)$ is related to the wave function $\psi (x,t)$ by means of
Eq.(\ref{eq28QL}). The equation (\ref{eq30}) and its 3D generalization were
proposed in \cite{Z15} to study wave propagation or kinetics in a nonlinear
media with fractal properties (see also \cite{Z16}, \cite{Z17}). Following
from Eq.(\ref{eq30}) a fractional generalization of the nonlinear
Ginzburg-Landau equation has been developed and square integrability of its
solution has been established as well in \cite{Z15}.

\subsection{Nonlinear Hilbert-Schr\"odinger equation}

For the case when $s=2$ substitution Eq.(\ref{eqA15C}) into Eqs.(\ref{eq13})
and (\ref{eq14}) yields a new nonlinear integro-differential equation

\begin{equation}
i\hbar \frac{\partial \psi (x,t)}{\partial t}=\omega \psi (x,t)-\pi V_0%
\mathcal{H}\left\{ \partial _x\psi (x,t)\right\} +\frac{\delta U(|\psi |)}{%
\delta \psi ^{*}(x,t)},\qquad  \label{eq31}
\end{equation}
where $\mathcal{H}$ is defined in Eq.(\ref{eq35C})

When $U(|\psi |)$ is given by Eq.(\ref{eq28L}) we find from Eq.(\ref{eq31})
the linear Hilbert-Schr\"odinger equation

\begin{equation}
i\hbar \frac{\partial \phi (x,t)}{\partial t}=-\pi V_0\mathcal{H}\left\{
\partial _x\phi (x,t)\right\} +U(x)\phi (x,t).  \label{eq29h}
\end{equation}
When $U(|\psi |)$ is given by Eq.(\ref{eq29L}) we obtain from Eq.(\ref{eq31}%
) the nonlinear Hilbert-Schr\"odinger equation 
\begin{equation}
i\hbar \frac{\partial \phi (x,t)}{\partial t}=-\pi V_0\mathcal{H}\left\{
\partial _x\phi (x,t)\right\} +U_0|\phi (x,t)|^2\phi (x,t).  \label{eq30h}
\end{equation}

Finally, we note that for $s>3$ it follows from Eq.(\ref{eqA17C}) that Eq.(%
\ref{eq25}) takes the form of the nonlinear Schr\"odinger-like equation with
extra term $\omega \psi (x,t)$ originated because of the energy parameter $%
\omega $

\begin{equation}
i\hbar \frac{\partial \psi (x,t)}{\partial t}=\omega \psi (x,t)-\left( \frac{%
V_0\zeta (s-2)}2\right) \partial _x^2\psi (x,t)+\frac{\delta U(|\psi |)}{%
\delta \psi ^{*}(x,t)},\qquad s>3,  \label{eq31}
\end{equation}
here $\zeta (s)$ is the Riemann zeta function.

\section{Quantum exciton propagator}

To get insight on impact of long-range interaction on 1D quantum dynamics
let us focus on the discrete linear problem associated with Eq.(\ref{eq5})

\begin{equation}
i\hbar \frac{\partial \psi _n(t)}{\partial t}=\varepsilon \psi
_n(t)-\sum\limits\Sb m  \\ (n\neq m)  \endSb V_{n-m}\psi _m(t),
\label{eq32q}
\end{equation}
with long-range potential $V_{n-m}$ given by Eq.(\ref{eq2C}).

Suppose that we know the solution $\psi _{n^{\prime }}(t^{\prime })$ of Eq.(%
\ref{eq32q}) at some time instant $t^{\prime }$ at the site $n^{^{\prime }}$%
. Then solution $\psi _n(t)$ at later time $t$, ($t>t^{\prime }$), and site $%
n$ will be

\begin{equation}
\psi _n(t)=\sum\limits_{n^{\prime }}G_{n,n^{\prime }}(t|t^{\prime })\psi
_{n^{\prime }}(t^{\prime }),  \label{eq333q}
\end{equation}
where $G_{n,n^{\prime }}(t|t^{\prime })$ is quantum exciton propagator, that
is probability of exciton transition from site $n^{\prime }$ at the moment $%
t^{\prime }$ to site $n$ at the moment $t$.

It follows from Eq.(\ref{eq32q}) and Eq.(\ref{eq333q}) that $G_{n,n^{\prime
}}(t|t^{\prime })$ is governed by the equation

\begin{equation}
i\hbar \frac{\partial G_{n,n^{\prime }}(t|t^{\prime })}{\partial t}%
=\varepsilon G_{n,n^{\prime }}(t|t^{\prime })-\sum\limits\Sb m  \\ (n\neq m) 
\endSb V_{n-m}G_{m,n^{\prime }}(t|t^{\prime }),\qquad t>t^{\prime },
\label{eq33q}
\end{equation}
with the initial condition $G_{n,n^{\prime }}(t|t)=\delta _{n,n^{\prime }}$,
here $\delta _{n,n^{\prime }}$ is the Kronecker symbol. Let us put for
simplicity $n^{\prime }=0$ and $t^{\prime }=0$ and introduce a notation 
\begin{equation}
G_n(t)\equiv G_{n,0}(t|0).  \label{eq34q}
\end{equation}
It yields

\begin{equation}
i\hbar \frac{\partial G_n(t)}{\partial t}=\varepsilon G_n(t)-\sum\limits\Sb %
m  \\ (n\neq m)  \endSb V_{n-m}G_n(t),\qquad t\geq 0,  \label{eq35q}
\end{equation}
with the initial condition

\begin{equation}
G_n(t)|_{t=0}=\delta _{n,0}.  \label{eq355q}
\end{equation}

To get the continuum version of Eq.(\ref{eq35q}) we apply transformations
similar to Eq.(\ref{eq4C}) and Eq.(\ref{eq5C}). That is we define $G(k,t)$ as

\begin{equation}
G(k,t)=\sum\limits_{n=-\infty }^\infty e^{-ikn}G_n(t),  \label{eq36q}
\end{equation}

\begin{equation}
G_n(t)=\frac 1{2\pi }\int\limits_{-\pi }^\pi dke^{ikn}G(k,t),  \label{eq361q}
\end{equation}
and $G(k,t)$ satisfies the equation

\begin{equation}
i\hbar \frac{\partial G(k,t)}{\partial t}=\left( \omega +\mathcal{V}%
(k)\right) G(k,t),\qquad t\geq 0,  \label{eq37q}
\end{equation}
where $\mathcal{V}(k)$ is the same as in Eq.(\ref{eq7C}) and the energy
parameter $\omega $ is defined by Eq.(\ref{eq8q}). The initial condition Eq.(%
\ref{eq355q}) now becomes

\begin{equation}
G(k,t=0)=1.  \label{eq371q}
\end{equation}
For further consideration it is convenient to introduce the quantum
propagator $g(k,t)$ which is

\begin{equation}
g(k,t)=\exp \{i\frac{\omega t}\hbar \}G(k,t).  \label{eq3711q}
\end{equation}
Thus, $G(k,t)$ can be expressed in terms of $g(k,t)$ as follow

\begin{equation}
G(k,t)=\exp \{-i\frac{\omega t}\hbar \}g(k,t).  \label{eq37111}
\end{equation}
It follows from Eq.(\ref{eq37q}) that the propagator $g(k,t)$ is governed by 
\begin{equation}
i\hbar \frac{\partial g(k,t)}{\partial t}=\mathcal{V}(k)g(k,t),\qquad t\geq
0,  \label{eq3712q}
\end{equation}
with the initial condition

\begin{equation}
g(k,t=0)=1.  \label{eq3713q}
\end{equation}
The solution of the problem Eqs.(\ref{eq3712q}) and (\ref{eq3713q}) is 
\begin{equation}
g(k,t)=\mathrm{exp}(i\frac{\mathcal{V}(k)t}\hbar ).  \label{eq372q}
\end{equation}
By substituting this solution into Eqs.(\ref{eq37111}) and (\ref{eq361q}) we
find

\begin{equation}
G_n(t)=\frac 1{(2\pi )}\int\limits_{-\pi }^\pi dk\exp (ikn-i\frac{(\omega +%
\mathcal{V}(k))t}\hbar ),  \label{eq39q}
\end{equation}
or, after extraction of the energy-time factor $\exp \{-i\frac{\omega t}%
\hbar \}$, 
\begin{equation}
G_n(t)=\exp \{-i\frac{\omega t}\hbar \}g_n(t),  \label{eq39Aq}
\end{equation}
where the lattice propagator $g_n(t)$ has been introduced by

\begin{equation}
g_n(t)=\frac 1{(2\pi )}\int\limits_{-\pi }^\pi dk\exp (ikn-i\frac{\mathcal{V}%
(k)t}\hbar ).  \label{eq399q}
\end{equation}

Generalization to 1D lattice quantum exciton propagator g$_{n,n^{\prime
}}(t|t^{\prime })$ which describes transition from site $n^{\prime }$ at the
time instant $t^{\prime }$ to site $n$ at the time instant $t$, is obvious

\begin{equation}
g_{n,n^{\prime }}(t|t^{\prime })=\frac 1{(2\pi )}\int\limits_{-\pi }^\pi
dk\exp \{ik(n-n^{\prime })-i\frac{\mathcal{V}(k)(t-t^{\prime })}\hbar \}.
\label{eq40q}
\end{equation}
This is the quantum exciton propagator which describes 1D transport discrete
in space and continuous in time. Let us note that g$_{n,n^{\prime
}}(t|t^{\prime })$ satisfies the following conditions:

1. Normalization condition

\begin{equation}
\sum\limits_{n=-\infty }^\infty g_{n,n^{\prime }}(t|t^{\prime })=1,
\label{eq41q}
\end{equation}

2. Consistency condition

\begin{equation}
g_{n,n^{\prime }}(t_1|t_2)=\sum\limits_{m=-\infty }^\infty
g_{n,m}(t_1|t^{\prime })\cdot g_{m,n^{\prime }}(t^{\prime }|t_2).
\label{eq42q}
\end{equation}

The last condition means that exciton propagator $g_{n,n^{\prime
}}(t|t^{\prime })$ can be considered as a transition quantum amplitude, and
for propagations occurring in succession in time transition amplitudes are
multiplied.

Further we will study behavior of g$_n(t)$ at large $|n|$, when the main
contribution to the integral Eq.(\ref{eq399q}) comes from small $k$.
Therefore, we expand integral over $k$ from $-\infty $ up to $\infty $

\begin{equation}
g_n(t)=\frac 1{2\pi }\int\limits_{-\infty }^\infty dk\exp (ikn-i\frac{\gamma
_sk^{\nu (s)}t}\hbar ),  \label{eq43q}
\end{equation}
with

\begin{equation}
\nu (s)=\{\QATOP{2,\ \qquad \mathrm{for}\ \ s>3,}{s-1,\qquad \mathrm{for}\
2<s<3,}  \label{eq44q}
\end{equation}
and

\begin{equation}
\gamma _s=\{\QATOP{\frac{V_0\zeta (s-2)}2,}{D_s,}\QATOP{\mathrm{for}\ s>3,}{%
\qquad \mathrm{for}\ 2<s<3,}  \label{eq45q}
\end{equation}
where $D_s$ is given by Eq.(\ref{eqA155C}).

Asymptotic behavior of the quantum 1D propagator g$_n(t)$ at large $|n|$
depends on the parameter $s$. Indeed, when $s>3$ Eq.(\ref{eq43q}) goes to

\begin{equation}
g_n(t)=\left( \hbar /2\pi iV_0\zeta (s-2)t\right) ^{1/2}\exp \{-\frac{\hbar
|n|^2}{2iV_0\zeta (s-2)t}\},\qquad s>3,  \label{eq47q}
\end{equation}
if we take into account Eqs.(\ref{eq44q}) and (\ref{eq45q}).

When $2<s<3$ the integral in right-hand side of Eq.(\ref{eq43q}) can be
expressed in terms of the Fox's $H$-function \cite{Laskin} and we have

\begin{equation}
g_n(\tau )=\frac 1{|n|(s-1)}H_{2,2}^{1,1}\left[ \left( \frac \hbar {iD_s\tau
}\right) ^{1/(s-1)}|n|\mid \left( \QATOP{(1,1/(s-1),(1,1/2)}{(1,1),(1,1/2)}%
\right) \right] ,\qquad  \label{eq46q}
\end{equation}
where $H_{2,2}^{1,1}$ is the Fox's function (for definition see, for example 
\cite{Laskin}, \cite{Mathai}). From other hand, in the long-wave limit for $%
\ 2<s<3$ the integral in Eq.(\ref{eq43q}) can be estimated as

\begin{equation}
g_n(t)\simeq \frac 1\pi \Gamma (s)\sin (\frac{\pi (s-1)}2)\left( \frac{iD_st}%
\hbar \right) ^{s/(s-1)}\cdot \frac 1{|n|^s},\qquad 2<s<3.  \label{eq49q}
\end{equation}
Thus, the long-wave asymptotics at large $|n|$ of the quantum exciton
propagator $g_n(t)$ exhibits the power-law behavior for $2<s<3$. Transition
from Gaussian-like behavior Eq.(\ref{eq47q}) to power-law decay Eq.(\ref
{eq49q}) is due to long-range interaction (second term in the right hand of
Eqs.(\ref{eq32q}), (\ref{eq35q})). This transition can be interpreted as
phase transition. In fact, for $s>3$ propagator g$_n(t)$ given by Eq.(\ref
{eq47q}) has correlation length (or characteristic scale) which can be
describe by

\[
\Delta n\simeq \left( \frac{2V_0\zeta (s-2)t}\hbar \right) ^{1/2}, 
\]
as far as in all above considerations we put the lattice constant $a$ equal
1 for simplicity. As it follows from Eq.(\ref{eq49q}), the correlation
length is infinite, that is the correlation length doesn't exist for
long-wave excitons in 1D lattice with $\ 2<s<3$ power-law interaction.

Finally, let us note that the above discussed 1D quantum lattice dynamics
for imaginary time is similar to the random walk model, that is similar to
CTRW \cite{Z9}. Indeed, if we put $it\rightarrow \tau $, Eq.(\ref{eq40q}) is
transformed to 
\begin{equation}
P_{n,n^{\prime }}(\tau |\tau ^{\prime })=\frac 1{(2\pi )}\int\limits_{-\pi
}^\pi dk\exp \{ik(n-n^{\prime })-\frac{\mathcal{V}(k)(\tau -\tau ^{\prime })}%
\hbar \},  \label{eq61p}
\end{equation}
where $P_{n,n^{\prime }}(\tau |\tau ^{\prime })$ is transition probability,
i.e. the probability to walk from site $n$ to site $n^{\prime }$ at the time
interval $\tau -\tau ^{\prime }$. It is easy to see that $P_{n,n^{\prime
}}(\tau |\tau ^{\prime })$ is normalized 
\begin{equation}
\sum\limits_{n=-\infty }^\infty P_{n,n^{\prime }}(\tau |\tau ^{\prime })=1,
\label{eq612p}
\end{equation}
and satisfies the discrete version of the Kolmogorov-Chapmen equation

\begin{equation}
P_{n,n^{\prime }}(\tau _1|\tau _2)=\sum\limits_{m=-\infty }^\infty
P_{n,m}(\tau _1|\tau ^{\prime })\cdot P_{m,n^{\prime }}(\tau ^{\prime }|\tau
_2).  \label{eq62p}
\end{equation}

That is we came to a continuous in time and discrete in space (1D lattice)
random walk model. The obtained random walk model exhibits a phase
transition from the Brownian random walk ($s>3$) with finite correlation
length to the symmetric $\alpha $-stable (with $\alpha =s-1,$ $2<s<3$) or
the L\'evy random process with an infinite correlation length.

In the case of continuum space the integral over $k$ Eq.(\ref{eq61p}) can be
expanded from $-\infty $ up to $\infty $ and we obtain transition
probability distribution function $P(x,\tau |x^{\prime },\tau ^{\prime })$ 
\begin{equation}
P(x,\tau |x^{\prime },\tau ^{\prime })=  \label{eq63p}
\end{equation}
\[
\frac 1{(2\pi )}\int\limits_{-\infty }^\infty dk\exp \left( ik(x-x^{\prime
})-\frac{\gamma _sk^{\nu (s)}(\tau -\tau ^{\prime })}\hbar \right) , 
\]
where $\nu (s)$ and $\gamma _s$ are defined by Eq.(\ref{eq44q}) and Eq.(\ref
{eq45q}) reciprocally. This probability distribution function exhibits
transition from Gaussian-like behavior at $s>3$ to power-law decay at $2<s<3$
because of the power-law long-range interaction.

\section{Conclusions}

We have presented analytical developments on classical and quantum 1D
lattice dynamics with fractional power-law long-range interaction. The
classical case has been treated in the framework of the well-known
Frenkel-Kontorova chain model. Quantum lattice dynamics is considered
following Davydov's approach to dynamics of molecular excitons. It has been
shown that the long-range power-law interaction leads, in general, to a
nonlocal integral term in the equation of motion if we go from discrete to
continues space. In some particular cases with non-integer power $s$ for
power-law interaction the integral term can be expressed through the
fractional order derivative. In other words nonlocality from the long-range
interaction reveals the dynamics in the form of space derivatives of
fractional order.

As useful examples of classical lattice dynamics in the continuum media
limit we were able to derive in a unified form fractional sine-Gordon and
wave-Hilbert nonlinear equations.

In the quantum case we obtained linear a fractional Schr\"odinger equation,
fractional nonlinear Schr\"odinger equation, linear Hilbert-Schr\"odinger
equation, and nonlinear Hilbert-Schr\"odinger equation.

This applied approach permits one to see, in an explicit form, the interplay
between second order and fractional order space derivatives both in
classical and quantum dynamics.

We have observed that there exists critical value for the power of the
long-range potential such that separates the cases of fractional equations
from the well-known nonlinear dynamical equations with integer order
derivatives.

The long-range interaction impact on quantum lattice propagators has been
studied. It was shown that the exciton propagator exhibits a transition for
the same critical value of the power-law exponent from the well-known
Gaussian-like behavior to a power-law space decay. This transition can be
treated as a phase transition accompanying by an infinite growth of the
exciton correlation length.

It has been shown that in the imaginary time domain 1D quantum lattice
dynamics with long-range interaction can be considered as random walk. The
corresponding equations (Eq.(\ref{eq61p}) and Eq.(\ref{eq63p}) ) for the
transition probability of this random walk have been obtained. The random
walk model also exhibits a phase transition from the Brownian random walk ($%
s>3$) with finite correlation length to the L\'evy random walk ($2<s<3$)
with an infinite correlation length.

Finally, let us note that this approach can be generalized to 2D and 3D
classical and quantum lattice dynamics.

\section{Acknowledgments}

This work was supported by the Office of Naval Research Grant No.
N00014-02-1-0056, and the NSF Grant No. DMS-0417800. G.M. thanks S. Flach
for useful discussions. We also thank M. Shlesinger for reading the paper
and useful comments.

\section{Appendix}

\subsection{Polylogarithm as a power series}

The polylogarithm $\mathrm{Li}_s(z)$ is defined by \cite{Lewin}-\cite{Erdely}

\begin{equation}
\mathrm{Li}_s(z)=\sum\limits_{n=1}^\infty \frac{z^n}{n^s}=\frac z{\Gamma
(s)}\int\limits_0^\infty dt\frac{t^{s-1}}{e^t-z},  \label{eqA1}
\end{equation}
here $s$ is real parameter and argument $z$ is the complex argument. It is
easy to see that for $z=1$ the polylogarithm $\mathrm{Li}_s(1)$ reduces to
the well-known Riemann zeta function $\zeta (s)$, 
\begin{equation}
\mathrm{Li}_s(1)=\zeta (s)=\sum\limits_{n=1}^\infty \frac 1{n^s}.
\label{eqA1a}
\end{equation}
It seems that the quantum statistical mechanics is the best known field
where the polylogarithm arises in natural way. Indeed, the integral of the
Bose-Einstein distribution is expressed in terms of a polylogarithm as,

\begin{equation}
\mathrm{Li}_s(z)=\frac 1{\Gamma (s)}\int\limits_0^\infty dt\frac{t^{s-1}}{%
e^t/z-1}.  \label{eqA2}
\end{equation}
In statistical mechanics handbooks the above integral is referred as a Bose
integral or a Bose-Einstein integral (see, for example \cite{Landau}).

Next, the integral of the Fermi-Dirac distribution is also expressed in term
of a polylogarithm,

\begin{equation}
\mathrm{Li}_s(-z)=-\frac 1{\Gamma (s)}\int\limits_0^\infty dt\frac{t^{s-1}}{%
e^t/z+1}.  \label{eqA3}
\end{equation}
This integral is referred as a Fermi-Dirac integral \cite{Landau}.

It is convenient to write the argument as $e^\mu $, that is consider $%
\mathrm{Li}_s(e^\mu )$. Then one can use the Hankel contour integral for the
polylogarithm \cite{Watson},

\begin{equation}
\mathrm{Li}_s(e^\mu )=-\frac{\Gamma (1-s)}{2\pi }\oint\limits_Hdt\frac{%
(-t)^{s-1}}{e^{t-\mu }-1},  \label{eqA4}
\end{equation}
where $H$ represent the Hankel contour and $s\neq 1,2,3,$..., and pole of
the integrand does not lie on non-negative real axis at the $t=\mu $. The
integrand has a cut along the real axis from zero to infinity, with the real
axis being on the lower half of the sheet ($\mathrm{Im}t\leq 0$). For the
case where $\mu $ is real and non-negative, contribution of the pole has to
be count,

\begin{equation}
\mathrm{Li}_s(e^\mu )=-\frac{\Gamma (1-s)}{2\pi }\oint\limits_Hdt\frac{%
(-t)^{s-1}}{e^{t-\mu }-1}+2\pi iR,  \label{eqA5}
\end{equation}
where $R$ is the residue of the pole

\begin{equation}
R=\frac{\Gamma (1-s)(-\mu )^{s-1}}{2\pi }.  \label{eqA6}
\end{equation}
To get power series (about $\mu =0$) representation of the polylogarithm $%
\mathrm{Li}_s(e^{-\mu })$ we apply the Mellin transform following Ref. \cite
{Robinson}

\begin{equation}
M_s(r)=\int\limits_0^\infty duu^{r-1}\mathrm{Li}_s(e^{-u})=\frac 1{\Gamma
(s)}\int\limits_0^\infty du\int\limits_0^\infty dt\frac{u^{r-1}t^{s-1}}{%
e^{t+u}-1}.  \label{eqA7}
\end{equation}
Then we observe that changing of variables $t=xy$, $u=x(1-y)$ separates the
integrals and we obtain

\begin{equation}
M_s(r)=\frac 1{\Gamma
(s)}\int\limits_0^1dyy^{r-1}(1-y)^{s-1}\int\limits_0^\infty dx\frac{x^{s+r-1}%
}{e^x-1}=\Gamma (r)\mathrm{Li}_{s+r}(1)=\Gamma (r)\zeta (s+r).  \label{eqA8}
\end{equation}
Now through the inverse Mellin transform we have

\begin{equation}
\mathrm{Li}_s(e^{-\mu })=\frac 1{2\pi i}\int\limits_{c-i\infty }^{c+i\infty
}dr\mu ^{-r}\Gamma (r)\zeta (s+r),  \label{eqA9}
\end{equation}
here $c$ is a constant to the right of the poles of the integrand. The path
of integration may be converted into a closed contour, and the poles of the
integrand are those of the gamma function $\Gamma (r)$ at $r=-l$ with
residues $(-1)^l/l!$ ($l=0,-1,-2,$...$)$, and of the Riemann zeta function $%
\varsigma (s+r)$ at $r=1-s$ with residue $+1$. Summing the residues yields,
for $|\mu |<2\pi $ and $s\neq 1,2,3,$....

\begin{equation}
\mathrm{Li}_s(e^{-\mu })=\Gamma (1-s)(\mu )^{s-1}+\sum\limits_{l=0}^\infty 
\frac{\zeta (s-l)}{l!}(-\mu )^l.  \label{eqA10}
\end{equation}
This equation gives us is get power series (about $\mu =0$) representation
of the polylogarithm $\mathrm{Li}_s(e^{-\mu })$.

Let's note that if the parameter $s$ is a positive integer $n$, both the $%
l=n-1$ term and the gamma function $\Gamma (1-n)$ become infinite, although
their sum does not. For integer $l>0$ we have

\begin{equation}
\stackunder{s\rightarrow l+1}{\lim }\left[ \frac{\zeta (s-l)}{l!}\mu
^l+\Gamma (1-s)(-\mu )^{s-1}\right] =\frac{\mu ^l}{l!}\left(
\sum\limits_{m=1}^l\frac 1m-\ln (-\mu )\right)  \label{eqA11}
\end{equation}
and for $l=0$,

\begin{equation}
\stackunder{s\rightarrow 1}{\lim }[\zeta (s)+\Gamma (1-s)(-\mu )^{s-1}]=-\ln
(-\mu ).  \label{eqA12}
\end{equation}
The power series representation Eq.(\ref{eqA10}) will be used to get Eqs.(%
\ref{eqA15})-(\ref{eqA17}).

\subsection{Properties of function $\mathcal{V}(k)$}

It is easy to see that $V(k)$ given by Eq.(\ref{eq8C}) with $V_n$ defined by
Eq.(\ref{eq2C}) can be expressed in terms of the polylogarithm $\mathrm{Li}%
_s(z)$

\begin{equation}
V(k)=2V_0\sum\limits_{n=1}^\infty \frac{\cos kn}{n^s}=2V_0\cdot \mathrm{%
Re\{Li}_s(e^{-ik})\}.  \label{eqA13}
\end{equation}
Further, the function $\mathcal{V}(k)$ defined by Eq.(\ref{eq7C}) becomes

\begin{equation}
\mathcal{V}(k)=2V_0\zeta (s)\mathrm{Re(1-}\frac{\mathrm{Li}_s(e^{-ik})}{%
\zeta (s)}).  \label{eqA14}
\end{equation}
Taking into account power series representation given by Eq.(\ref{eqA10}) we
find from Eq.(\ref{eqA14}) at the limit $k\rightarrow 0$ that

\begin{equation}
\mathcal{V}(k)\sim \frac{V_0\pi }{\Gamma (s)\sin (\frac{\pi (s-1)}2)}%
|k|^{s-1},\quad \quad 2\leq s<3,  \label{eqA15}
\end{equation}

\begin{equation}
\mathcal{V}(k)\sim -\frac{V_0k^2}2\ln k^2,\quad s=3,  \label{eqA16}
\end{equation}
\begin{equation}
\mathcal{V}(k)\sim \frac{V_0\zeta (s-2)}2k^2,\quad s>3.  \label{eqA17}
\end{equation}

\end{document}